\documentclass[aps,floatfix, twocolumn]{revtex4} 

\usepackage{graphicx}
\usepackage{dcolumn}
\usepackage{bm}
\usepackage{longtable}
 \usepackage{amsmath}
 \usepackage{amssymb}
 \usepackage{eurosym}
 \usepackage{subfigure}
\usepackage{color}
\usepackage{float}
\graphicspath{{./images/}}

\begin{document}

\title{Dynamical origins of the community structure of multi-layer societies}

\author{Peter Klimek$^{1,\dagger}$, Marina Diakonova$^{2,\dagger}$, V\'ictor M. Egu\'iluz$^{2}$, Maxi San Miguel$^{2}$, Stefan Thurner$^{1,3,4}$}
\email{stefan.thurner@meduniwien.ac.at}
\affiliation{$^1$Section for Science of Complex Systems; CeMSIIS; Medical University of Vienna; 
Spitalgasse 23; A-1090; Austria\\ 
$^2$ Instituto de F\'isica Interdisciplinar y Sistemas Complejos IFISC (CSIC-UIB); E07122 Palma de Mallorca; Spain\\
$^3$IIASA, Schlossplatz 1, A 2361 Laxenburg; Austria\\
$^4$Santa Fe Institute; 1399 Hyde Park Road; Santa Fe; NM 87501; United States\\
$^{\dagger}$ These authors contributed equally to this work.
}

\begin{abstract} 
Social structures emerge as a result of individuals 
managing a variety of different of social relationships.
Societies can be represented as highly structured dynamic multiplex networks. 
Here we study the dynamical origins of the specific community structures of a large-scale social multiplex network of a human society 
that interacts in a virtual world of a massive multiplayer online game.
There we find substantial differences in the community structures of different social actions, represented by the various network layers in the multiplex.
Community size distributions are either similar to a power-law or appear to be centered around a size of 50 individuals.
To understand these observations we propose a voter model that is built around the principle of triadic closure. It explicitly models 
the co-evolution of node- and link-dynamics across different layers of the multiplex.
Depending on link- and node fluctuation rates, the model exhibits an anomalous shattered fragmentation transition, 
where one layer fragments from one large component into many small components. 
The observed community size distributions are in good agreement with the predicted fragmentation in the model.
We show that  the empirical pairwise similarities of network layers, in terms of link overlap and degree correlations, practically coincide with the model.
This suggests that several detailed features of the fragmentation in societies can be traced back to the triadic closure processes.
 \end{abstract}

\maketitle
\section{Introduction} 

Societies are organized dynamical patterns that emerge from the social actions of individuals.
These arrange an array of different types of social relationships (e.g. friendship, marriage, co-workers, ...) to form 
stable groups, organizations, or institutions of various sizes.
Each type of social relation defines a social network of its own.
These networks are not independent of each other but co-evolve with the other networks in the society.
Societies can be understood as the collection of these networks and can be represented as a dynamic, 
co-evolving multiplex network, i.e. a network where a set of nodes can be connected by links of more than 
one type \cite{Szell2010,Boccaletti14, Kivela14}.
The topological structures of these network layers can vary dramatically, depending on the type of the corresponding social interactions.
For example, it has been shown that network layers corresponding to cooperative behavior can be characterized by high clustering, 
high reciprocity, and high link overlap, whereas layers encoding aggressive behavior exhibit pronounced power-law degree distributions \cite{Szell2010, SzellSN10}.
It has also been shown that many characteristics of the individual multiplex layers, such as their degree distributions, clustering coefficients, or the probabilities for nodes to acquire new links, can be understood from the assumption that the link dynamics in networks is driven by the process 
of triadic closure (i.e. the tendency that nodes with common neighbors will establish links between themselves) 
together with a finite lifetime of links (i.e. the typical rate at they are added to and removed from the network) \cite{Davidsen02, Thurner2013}.
Another generic feature of social networks is that individuals tend to form communities, i.e. groups of nodes that share more links with each other than with nodes outside of the community \cite{Girvan02, Fortunato10}.
There is evidence that the organization of community structure in human societies follows principles that are deeply rooted in human psychology, such as a hierarchically nested organization of communities of different sizes \cite{Dunbar2003,Fuchs14} or Dunbar's number \cite{Dunbar1993}, a hypothesized upper cognitive limit to the number of people with whom humans can share stable social relationships.
   
In this work we investigate the dynamical origins of the community structure of societies in multiplex networks.
We study how the different layers in the multiplex influence each other and investigate the resulting consequences of 
these interactions for the community structures in the individual layers. 
Community structure will be simply characterized by the community size distributions in the different network layers, 
i.e. the probability for an individual (node) to be part of a community of a given size in a particular layer.
As a data set we use the comprehensive dynamic social multiplex network of the {\it Pardus} society 
\cite{Szell2010, SzellSN10, Szell12, Thurner2013, ThurnerConduct2012, Szell2013}.
This is a virtual society of more than 380,000 players with different social and economic interactions 
taking place in the open-ended massive multiplayer online game {\it Pardus}.  
We find a substantial amount of heterogeneity in the community size distributions across the different layers. 

In this work we want to understand if the empirical observations can be reminiscent of a so-called fragmentation transition.
Fragmentation is the phenomenon in which a network might undergo a transition from a state 
where almost all nodes belong to a single giant component to a {\em fragmented} state in which the network 
breaks into many smaller components.
To this end we propose a new type of voter model (VM). VMs  introduced in \cite{Liggett1975, Vazquez2008NJP}
have repeatedly been shown to exhibit  fragmentation transitions \cite{Gross2008, Vazquez2013}.
In the co-evolutionary VM (CVM) \cite{Vazquez2008} a node can either change its internal state to that of its neighbor or rewires one of its existing links towards 
a node that has the same internal state.
Fragmentation transitions have been shown to be generic features of rewiring processes and largely 
independent of the underlying evolution rules for the internal states \cite{Zimmermann2001,Zimmermann2004}.
CVMs of this kind account for a wide class of phenomena ranging from opinion-formation 
\cite{Eguiluz2005, Vazquez2007, Couzin2011, Zschaler2012, Adrian2014} 
to speciation in ecosystems \cite{Dieckmann1999,Cantor2013}.
In an extension of the CVM to multiplex networks (MCVM) nodes have the same internal states in each 
layer and it is assumed that the rewiring dynamics takes place on a different time-scale in 
each of the layers \cite{Diakonova2014}. This MCVM has an anomalous transition called {\em shattered fragmentation} 
in which one layer assumes a fragmented state whose topological properties would be different under the same model parameters in a single layer CVM \cite{Diakonova2014}.
It has been shown that the driving force behind this transition is the asymmetry between rewiring rates in 
different network layers, i.e. the individual lifetime of links in each of the layers \cite{Diakonova2014}.

To reconcile the dynamics of the CVM with the empirical dominance of triadic closure, where a substantial 
amount of newly created links in social networks connect nodes that already share common neighbors 
\cite{Simmel08, Heider46, SzellSN10, Thurner2013}, 
we propose a novel type of VM in which the rewiring step is carried out by a triadic closure process, and call it the 
triadic closure voter model (TCVM).
We study the dynamics of the TCVM on a single-layer and on a multiplex network (MTCVM). 
We show that the MTCVM displays a novel fragmentation transition that is again different from  shattered fragmentation in the MCVM.
We investigate whether the fragmentation behavior of the MTCVM is compatible with the community structure observed in  {\it Pardus} multiplex network data. 
To understand the extent to which the microscopic dynamics of the model multiplex is compatible with the data, 
we compare results for pairwise similarity measures of layers in terms of the properties of nodes (states) and links (overlap and degree).

\section{Data and Methods} 
\label{sec:DandM}

\subsection{Data} 

The Pardus dataset contains all actions of more than 380,000 players in a massive multiplayer online game.
The players interact in an virtual, open-ended game universe to connect with other players to achieve self-posed 
goals, such as accumulating wealth and influence. Players can engage in three different types of cooperative interactions.
They can establish mutual friendship links, exchange private, one-to-one messages and trade with each other in the game.
The data can be represented as a dynamic multiplex network $M^{\alpha}(t)$ where the index $\alpha$ labels the 
adjacency matrix of the network given by interactions of type $\alpha$ at time $t$.
The multiplex $M^{\alpha}(t)$ is constructed for each month (30 days) over one year of data, from Sep 2007 to Sep 2008.
Two players are linked in a corresponding multiplex layer if they had a friendship link ($\alpha = \textrm{friend}$), traded with 
each other ($\alpha = \textrm{trade}$), or exchanged a private message ($\alpha = \textrm{communication}$) within a given month.
For a particular $t$ we include all players that have at least one link in each of the multiplex layers.
For more information on the topology and structure of the Pardus multiplex network see \cite{Szell2010, SzellSN10}.
Per construction, each layer has the same average number of nodes $N=3.1(0.2) \cdot 10^3$. Numbers in brackets denote standard deviations.
Table \ref{data} shows results for the average number of links in each layer, $L^{\alpha}$.
The trade network shows the highest link density,  the friendship and communication network have similar numbers of links.

It has been shown that for the given time-span the three network layers $\alpha$ are in a 
stationary state in the sense that links are added and removed with comparable rates. These rates 
are orders of magnitude larger than the rates at which nodes are added or removed, see \cite{Thurner2013}.
The dynamics within the network layers is therefore dominated by rewiring processes.
The rewiring rate $p_{\alpha}$ is defined as the average value of the probabilities that a link will be 
added or removed, that is rewired, in layer $\alpha$. Table \ref{data} shows results for the rewiring rates $p_{\alpha}$ in the three layers.
$p_{\alpha}$ in the friendship network is orders of magnitude smaller than in the communication and trade networks.
This can be understood by the difference in processes that govern the interactions in these layers.
In the friendship network a link persists after it has been formed until the link is removed or one of the players leaves the game.
In the message and communication network, on the other hand, a link is only formed between two players 
if at least one interaction took place within the considered time interval.
This results in a substantially lower turnover of links in the friendship network than in the other layers.
We will therefore refer to the friendship layer as being {\em slow} and to the trade and message layers as being {\em fast} 
in terms of the average time between two consecutive rewiring events in the given layer. 
Note that although friendship links have the longest survival time (i.e. lowest turnover $p_{\alpha}$), the friendship network 
has also the smallest number of links $L^{\alpha}$.

\begin{table}[tbp]
\caption{Overview of characteristics of the Pardus multiplex network layers. For each layer $\alpha$ we show the values of the rewiring 
probability $p_{\alpha}$ and the average number of links $L^{\alpha}$. The rewiring probability $p_{\alpha}$ is orders of magnitude 
smaller for the friendship network, compared to communication and trade. Each layer has the same average number of nodes, $N=3.1(0.2) \cdot 10^3$.}
\label{data}
\begin{tabular}{l l l l}
$\alpha$ & friendship & trade & communication \\
\hline
$p_{\alpha}$ & $0.004(1)$ & $0.27(1)$ & $0.35(2)$\\
$L^{\alpha}$ & $1.5(0.2) \cdot 10^4$ & $5.6(0.2) \cdot 10^4$ & $1.9(0.3) \cdot 10^4$\\
\end{tabular}
\end{table}

\subsection{Community detection}

From the network layers $M^{\alpha}(t)$, for data and model, we identify the communities $C^{\alpha}(t) = \{C^{\alpha}(1,t), C^{\alpha}(2,t),  \dots, C^{\alpha}({N_c^{\alpha},t)}  \}$, where the $i$-th community $C^{\alpha}(i,t)$ is the set with a number of $n_i^{\alpha}(t)$ nodes within community $i$ at time $t$.
We use the OSLOM community detection algorithm \cite{Lancichinetti2011}, 
which retrieves only statistically significant communities and which does not suffer from the so-called {\em resolution limit} problem (failure of the detection of small communities) of other approaches \cite{Fortunato07}.
It also allows to detect the {\em absence} of community structure as well as {\em homeless} nodes, that do not belong to any community.
We use the OSLOM implementation as provided by the authors for unweighted networks with a coverage parameter 
of $1$ and a standard significance threshold of $p<0.1$, \cite{Lancichinetti2011}.

\subsection{Components in the model}

To understand the fragmentation behavior of the TCVM we describe the organization of the model-networks 
into {\em components} by the following observables. 
$N_c^{\alpha}(t)$ is the number of components in network layer $\alpha$ at time $t$.
We will refer to the time average of $N_c^{\alpha}(t)$ over $T$ consecutive time-steps by dropping the dependence on $t$, i.e. $N_c^{\alpha} \equiv \tfrac{1}{T} \sum_t N_c^{\alpha}(t)$.
The size of the $k$-largest component at $t$ is denoted $S_k^{\alpha}(t)$.
The time average of the $k$-largest component size, $S_k^{\alpha}$, is again denoted by dropping the dependence on $t$.

\subsection{Similarity measures}
 
The similarity of the sets of links in two layers, $M^{\alpha}(t)$ and $M^{\beta}(t)$, is measured by the 
Jaccard coefficient $J(\alpha, \beta, t)$. 
Let $E^{\alpha (\beta)}(t)$ be the set of links in $M^{\alpha (\beta)}(t)$.
The Jaccard coefficient is given by $J(\alpha, \beta, t) = \tfrac{|E^{\alpha}(t) \cup E^{\beta}(t)|}{|E^{\alpha}(t) \cap E^{\beta}(t)|}$.
Let us further denote the degree sequence of $M^{\alpha (\beta)}(t)$ by $k^{\alpha (\beta)}(t)$ and the sequence of the 
ranks of the degrees by $Rk(k^{\alpha (\beta)})(t)$.
The degree correlation, $\rho(k^{\alpha}(t), k^{\beta}(t))$, is Pearson's correlation coefficient between the degree sequences 
$k^{\alpha}(t)$ and $k^{\beta}(t)$. Similarly, the degree rank correlation, $\rho(Rk(k^{\alpha}(t)), Rk(k^{\beta}(t)))$, is 
Pearson's correlation coefficient between the degree rank sequences $Rk(k^{\alpha}(t))$ and $Rk(k^{\beta}(t))$.
As before, we refer to the time averages of $J(\alpha, \beta, t)$, $\rho(k^{\alpha}(t), k^{\beta}(t))$, and $\rho(Rk(k^{\alpha}(t)), Rk(k^{\beta}(t)))$, 
by dropping the time dependence in the variables, respectively.

\section{Results}

\subsection{Community structure in the {\em Pardus} multiplex}

\begin{figure}[tbp]
\begin{center}
 \includegraphics[width = 0.49\textwidth, keepaspectratio = true]{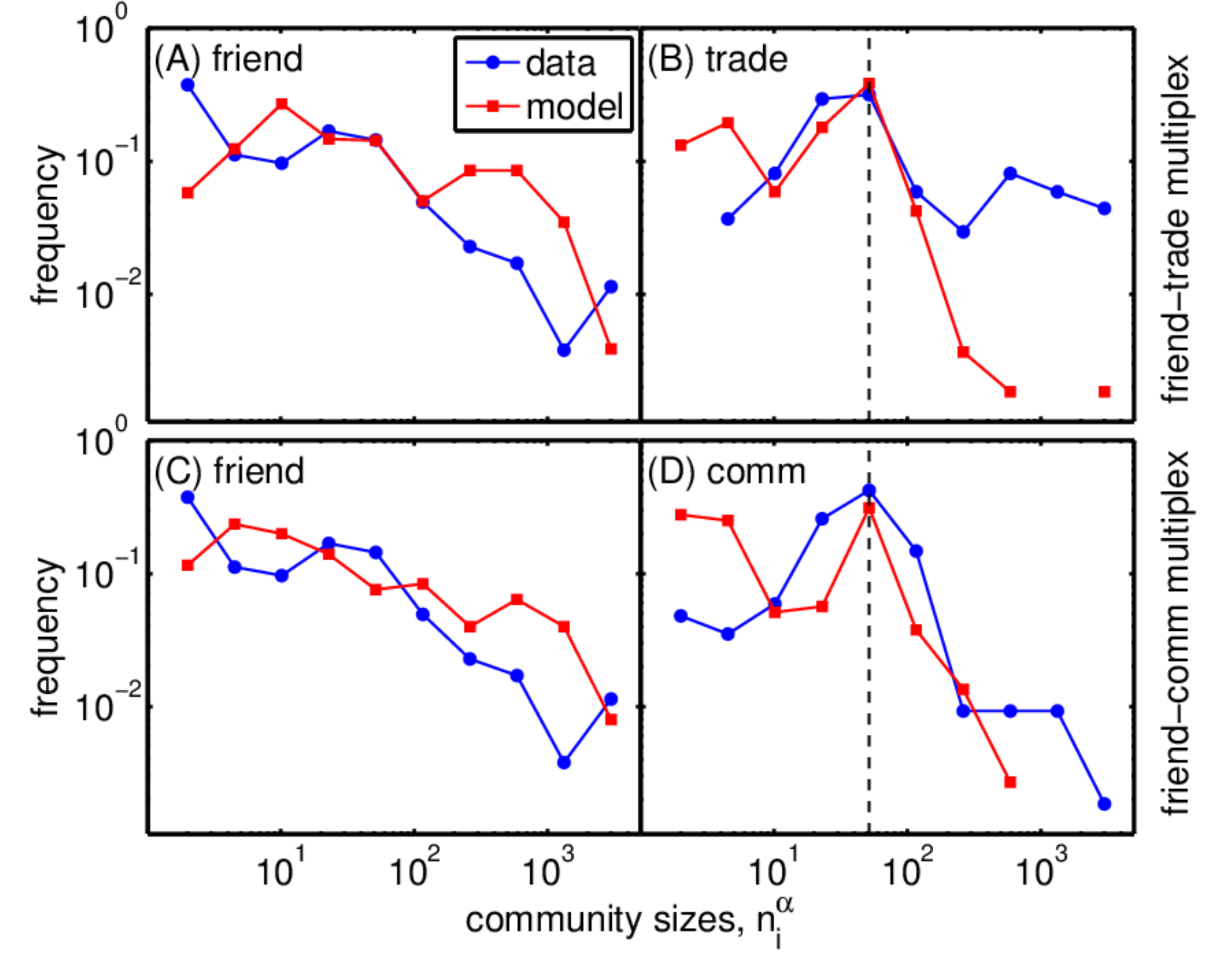}
\end{center}
 \caption{Community size distributions for the {\em Pardus} multiplex data (blue dots) for the friendship layer (A and C), the trade layer (B), and the communication layer (D). The friendship layer shows a decrease of frequencies of communities show as a function of their size, whereas the trade and message layers show an additional peak at about 50 (dotted lines in B and D). These results are compared to the community size distributions obtained from the model (red squares). For this comparison the three-layer multiplex from the data is decomposed into two two-layer multiplex networks, the friendship-trade multiplex (top row) and the friendship-communication multiplex (bottom row). Model community size distributions for the friendship layer are compatible with the data. Note that the model shows a distinct peak for the trade and communication networks that coincides with the peaks observed in the data.}
 \label{empirics}
\end{figure}

The time-averaged distribution functions for the community sizes in the three network layers are shown in figs. \ref{empirics}(A)-(D).
There is a clear discrepancy between the distribution observed in the {\em slow} friendship layer when compared to the trade and 
message networks, that are characterized by substantially larger rewiring rates.
For the friendship layer the frequencies of community sizes are clearly a decreasing function in size (roughly following a power-law), 
whereas there exist distinct peaks in the distributions of community sizes in the trade and message layers. 
In both {\em fast} layers these peaks are centered around a community size of 50.

\subsection{The Triadic Closure Voter Model (TCVM)} 
\label{sec:the_triadic_closure_model}

The standard binary-state CVM on a single network as introduced in \cite{Vazquez2008, Vazquez2008NJP} is given as follows.
Each node, $i$, is described as a time-dependent, binary, internal state $\sigma_i(t) \in \{0,1\}$ subject to the following update rule.
(i)  Pick node $i$ and one of its neighbors, $j$, at random.
If the internal states of these nodes differ, $\sigma_i(t) \neq \sigma_j(t)$, then (iia) with probability $p$ the link between $i$ and $j$ 
is removed and a link between $i$ and a different node $k$ is formed. $k$ is randomly chosen from the set of all nodes 
disconnected to, but in the same internal state as node $i$, $\sigma_i(t) = \sigma_k(t)$.
If no such node $k$ exists, the link is not re-wired.
Otherwise, (iib) with probability $1-p$, the state of node $i$ is changed to $\sigma_j(t)$.
This model has one parameter, the rewiring rate $p$, which defines the preference of a node to re-wire the link over changing its state.

We now introduce the triadic closure voter model (TCVM) on a single network.
The TCVM is motivated by the empirical fact that links that connect nodes that share 
common neighbors are more likely to form than links that do not connect such nodes, 
i.e. the process of triadic closure \cite{Simmel08, Heider46, Thurner2013}. 
In the TCVM the rewiring step (iib) of the CVM is replaced by the following update rule.
With the triadic closure probability, $p_{tc}$, the new link is made between node $i$ and a different node $l$ that is randomly chosen from the set of all nodes that share at least one common neighbor with $i$ (but there is no connection between $i$ and $l$, yet).
If no such node $l$ exists, no rewiring takes place.
Otherwise, with probability $1-p_{tc}$, we follow the rewiring rule from step (iib) of the CVM.

The network is initialized as a random graph with size $N$ and average degree 
$\mu=4$, which results in a network that is initially connected. The initial distribution of internal states is random, 
so that each node has the same probability $1/2$ to be in one of the two possible states. 
Results for the dynamics and phase diagrams of the TCVM on single networks are discussed in the supplementary information, 
where it is shown that the system undergoes a fragmentation transition at a critical rewiring rate $p_c$.
For $p < p_c$, the network freezes into one giant component with all nodes in the same state, i.e. the system reaches consensus. 
For $p > p_c$, the network splits into two components of roughly equal size, but differing internal states 
(one component is composed of nodes with $\sigma_i(t)=0$,  the other with $\sigma_i(t)=1$). 
In this regime the internal states initially present in the system are preserved, but those who hold them become segregated from each other. 
Such fragmentation at high rewiring rates is typical for a range of dynamical and evolution rules \cite{Zimmermann2004}.
For the single layer case, this fragmentation transition is largely independent of the triadic closure probability, $p_{tc}$.
In the following we set $p_{tc} = 1$.

\begin{figure}
\centering
\includegraphics[width = 0.49\textwidth, keepaspectratio = true]{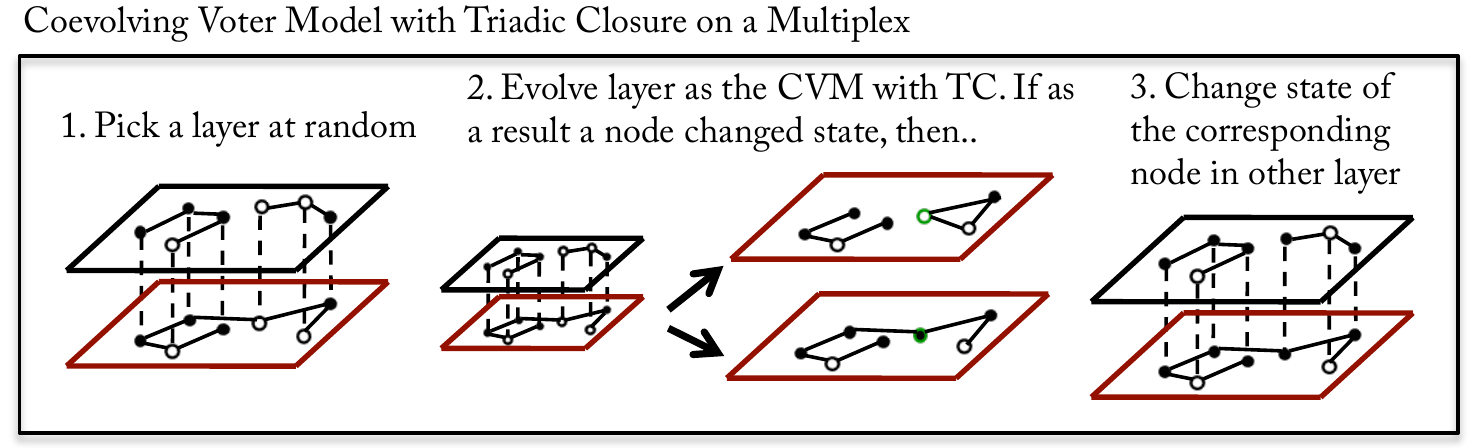}
\caption[]{Schematic representation of the coevolving voter model with triadic closure on a multiplex (MTCVM). We pick a layer at random (step 1) and evolve it according to the dynamics of the CVM with triadic closure (step 2). As a result the state of a node may change in both layers (step 3).}
\label{modelschema}
\end{figure}

\subsection{The TCVM on multiplex networks (MTCVM)}
We now study the TCVM on a multiplex network, the MTCVM, by joining two networks that both 
follow the update rules of the TCVM. The multiplex evolves by, first, picking a layer randomly and, 
secondly, by evolving that layer by using the update rules from the TCVM, see fig.~\ref{modelschema}.
This introduces a co-evolution of the two layers because the formation of new links depends on the 
internal state of nodes which is the same (!) in both layers for each of the nodes. 
Each of the layers thus has its own re-wiring probability, $p_1$ and $p_2$, respectively.

\begin{figure}
	\centering
\hspace*{-0.2 in}
\subfigure[$S_1^1/N$]{
\includegraphics[width = 0.23\textwidth, keepaspectratio = true]{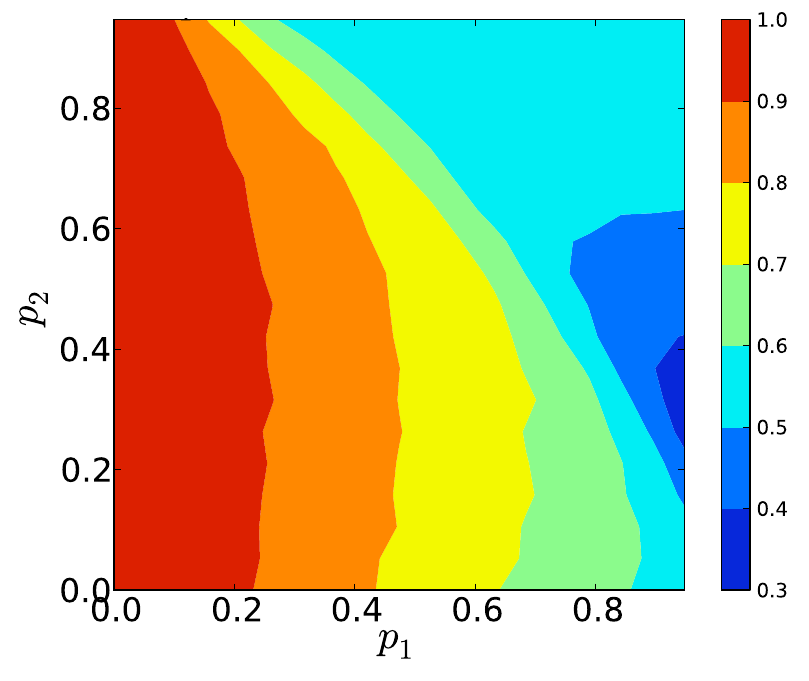}
\label{mult:a}
}%
\hspace*{-0.1 in}
\subfigure[$N_c^1/N$]{
\includegraphics[width = 0.23\textwidth, keepaspectratio = true]{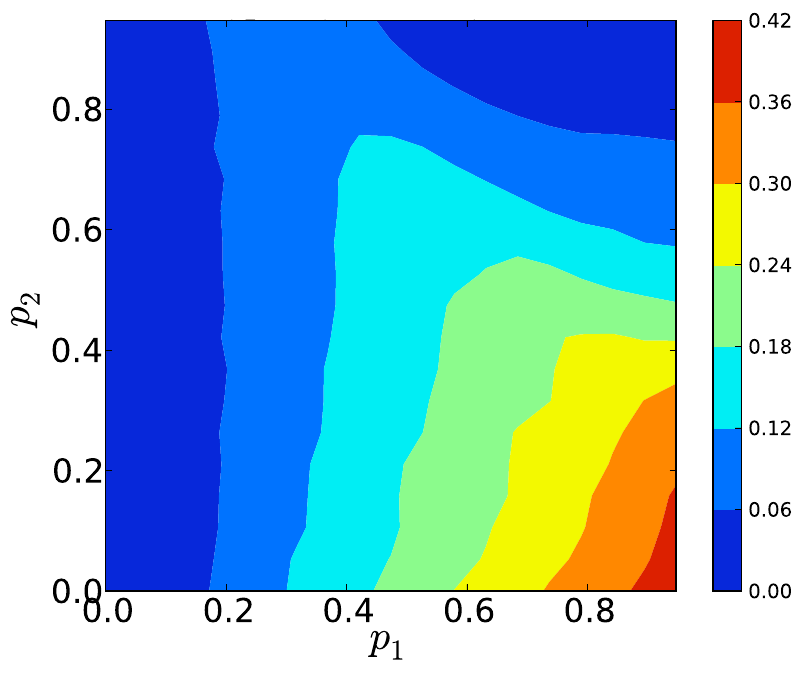}
\label{mult:c}
}
\caption[]{Shattered fragmentation in the MTCVM. Phase diagrams for (a)  fraction of nodes in the largest component, $S_1^1$, and (b)  
number of components $N_c^1$ relative to the system size $N$ are shown as a function of the re-wiring probabilities $p_1$ and $p_2$, 
respectively. The system undergoes a fragmentation transition where, with increasing $p_1$, layer 1 splits into a large number of small 
components.}
\label{mult}
\end{figure}

The MTCVM displays a novel fragmentation transition that is visible in the phase diagrams of the observables 
$S_1^{\alpha}$ (size of largest component) and $N_c^{\alpha}$ (number of components).
Fig. \ref{mult} shows the phase diagrams of (a) $S_1^1$ and (b) $N_c^1$ in layer 1 of the TCVM as a function of the rewiring 
probabilities $p_1$ and $p_2$. The results have been averaged over $10^3$ realizations of the final (absorbing) 
configurations of networks of size $N=250$.
Consider the case where layer 1 is connected to a static layer with $p_2  = 0$.
With increasing $p_1$ the largest component of layer 1 shrinks to a value around $0.5$, similar to the single-layer case.
$N_c$, on the other hand, increases by increasing $p_1$ in the same way as $S_1$ decreases.
This means that with increasing $p_1$ nodes leave the largest component and form small or 
isolated components of their own (and not a second, large component), i.e. $N_c$ increases.
This process has been called {\em shattered fragmentation} \cite{Diakonova2014}.
From the phase diagrams in fig. \ref{mult} it follows  by symmetry that in the static layer 2 the nodes 
remain in one giant component as layer 1 undergoes this shattered fragmentation.

Note that this fragmentation transition in the MTCVM is {\em different} from the fragmentation transition of the MCVM.
The MCVM can be recovered by setting $p_{tc} = 0$ in both layers and by introducing an additional parameter, the multiplexity $q$.
In the MCVM both layers contain $N$ nodes, but only a fraction $qN$ nodes exist in both layers, i.e. the internal state of the 
remaining $(1-q)N$ nodes depends only on the dynamics in their own layer.
The MCVM also displays a shattered fragmentation transition that is, however, only encountered below a critical value of $q$, $q<q_c<1$.
Under triadic closure, partial multiplexing (i.e. $q<1$) is no longer required to see shattered fragmentation.

\subsection{The MTCVM versus {\em Pardus} data}

One of the challenges when comparing the theoretical results with data is that the data contains 
three network layers, whereas, for simplicity, the model deals with only two layers.
We therefore decompose the three-layer multiplex network into two two-layer networks.
The most reasonable choices for this decomposition is to consider, both, the friendship-trade and the friendship-message multiplex networks.
Both of these two-layer multiplexes consist then of a {\em fast} and a {\em slow} layer in terms of rewiring rates $p_{\alpha}$.
We further assume that the model internal states, $\sigma_i(t)$, encode a hidden propensity of the players to interact with each other.
That is, upon meeting two players $i$ and $j$ will be more likely to cooperate with each other if they have the same internal and not directly observable states.

The community size distributions obtained from the model agrees well with the data.
In fig. \ref{empirics} we show the frequency of community sizes observed from 1000 realizations 
using $N=3100$ and the rewiring rates $p_{\alpha}$ of the two layers as measured for the respective 
layers in the data and compare these results to the data.
Results are shown for the friendship-trade multiplex (top row), namely (A) friendship and (B) trade, 
and for the friendship-communication multiplex (bottom row), (C) friendship and (D) communication.
In both model multiplexes the frequencies of community sizes decreases as a function of their size in 
strong resemblance to results observed in the data.
For the trade and communication layers, respectively, we also observe a clear peak in the frequencies of 
community sizes that coincides with the peaks observed in the data around a community size of 50.
Note that here we compare communities in the data to communities in the model (and not to components).
Clearly, the fragmentation behavior into components in the model is closely related to its community structure.
Given the rewiring probabilities $p_{\alpha}$ in the data, we would expect from the MTCVM that the friendship 
layer ($p^{friend}=0.004(1)$) shows one large component whereas the trade ($p^{trade}=0.27(1)$) and communication 
($p^{comm}=0.35(2)$) layers display fragmentation.
Indeed, we observe for the {\em communities} in, both, data and model for the friendship layer a roughly  
power-law-like size distribution with a small number of very large communities, whereas the trade and message 
layers fragment into a large number of smaller communities with a peak centered at around 50.
Data and model show the same behavior in terms of fragmentation into communities of various sizes.

\begin{figure}[tbp]
\begin{center}
 \includegraphics[width=70mm]{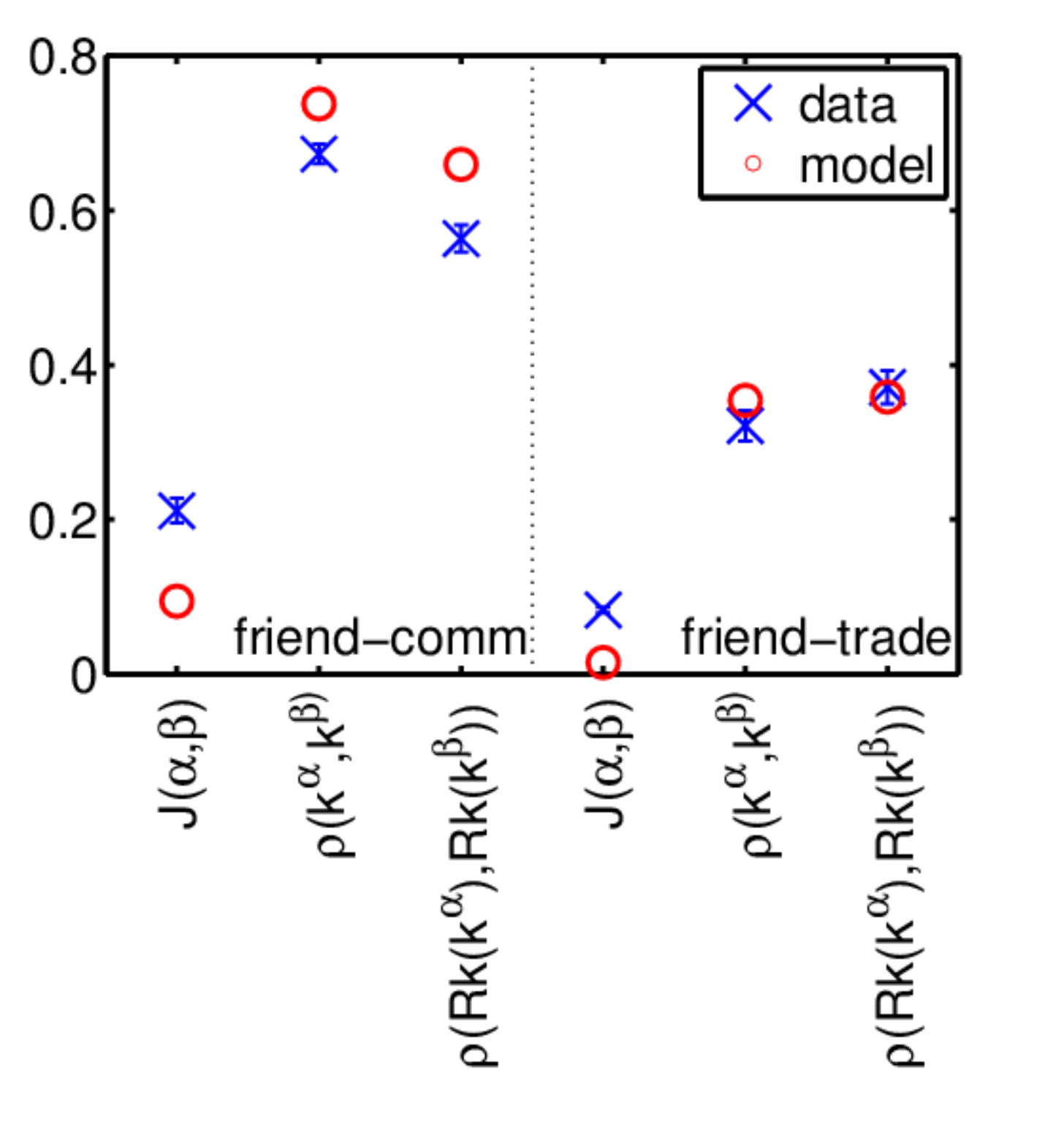}
\end{center}
 \caption{Similarity between pairs of multiplex layers for data (blue) and model (red). 
 We show the Jaccard coefficient of the edge sets of the friendship-communication multiplex (left) and the 
 friendship-trade multiplex (right), together with the degree correlation $\rho(k_{\alpha},k_{\beta})$ and the 
 degree rank correlation $\rho(Rk(k_{\alpha}),Rk(k_{\beta}))$. The model follows the similarities found  
 in the data, with the overlap between data and model being higher in the friendship-trade multiplex than the 
 friendship-communication multiplex.}
 \label{similarity}
\end{figure}

To study whether the {\em dynamics} of the MTCVM is able to describe the observed similarities between pairs 
of network layers in the data, we introduce a calibrated version of the MTCVM. We model the friendship-trade and 
the friend-communication multiplexes with corresponding $p_{\alpha}$ and initial conditions that are given by snapshots 
of the respective data layers at $t_1 = 100d$. For the friendship-communication multiplex the absorbing state was typically 
reached after $t \sim 10^5$, the friendship-trade multiplex did not do so within any reasonable time, 
i.e. for at least several orders of magnitude of simulation time. 
Therefore the friendship-trade multiplex was compared to snapshots taken at times ranging over 
several orders of magnitude, namely at simulation times $t=10^3, 10^4, 10^5$. 
In the data we found substantially higher levels of similarity in terms of the Jaccard coefficients, degree correlations, and rank degree 
correlations than in the model. This can be understood by the fact that the MTCVM does not contain any mechanism that 
explicitly increases the similarity of layers, such as by copying one link from one layer to the other. 
Such an inter-layer link correlation mechanism can be easily added to the MTCVM: 
In the rewiring step  a probability $p_{\text{sim}}$ is introduced to re-wire the 
link to another node in the same state to which the node is already connected in the {\em other} layer. 
We find that by introducing even a small modification of this type $(p_{\text{sim}} = 0.1)$ we are able to account for 
the observed values of the Jaccard similarity, degree correlations, and degree rank correlations, which are shown in 
fig. \ref{similarity} for data and model. Although the friendship-trade multiplex has not reached its absorbing state in these 
simulations, the variance of the similarity measures shows that they do not considerably change over time. 
The model exhibits the same trends as seen in the data, with the overlap between data and model being higher 
in the friend-trade multiplex than the friend-message multiplex, in particular for the degree correlation and the degree rank correlation.
This shows that the modified co-evolutionary dynamics of the MTCVM preserves the observed similarities between pairs of layers 
already for a very small probability $p_{\text{sim}}$ to ``copy'' a link from one layer to the other one.
We have confirmed that this value of $p_{\text{sim}}$ is indeed small enough that the community size distributions are unaltered by this inter-layer link correlation modification.
This link-copying mechanism has also no substantial impact on the phase diagrams of the model, since the mechanism is similar to a random linking process (i.e. a triadic closure probability, $p_{tc}$, smaller than one).
As already discussed, the phase diagrams are largely independent of such small variations of $p_{tc}$.

\section{Discussion}

We investigated the dynamical origins of the community structure of societies represented as dynamical multiplex networks.
In empirical data from the large-scale online game society {\em Pardus} we observed substantial differences in the community 
structures of individual network layers of this multiplex. 
While one layer is characterized by a small number of large communities and a power-law-like distribution of community sizes, 
in the other layers we find a peak of communities of intermediate size around 50.
We found that the time-scales on which the link dynamics takes place in the various network layers differ by several orders of magnitude.
Remarkably, we find that the power-law like distribution of community sizes is found in a layer with a very small re-wiring rate, 
whereas layers with the centered distribution are characterized by rates that are orders of magnitude higher.

To understand these empirical findings we proposed a generalization of the  co-evolutionary voter model on multiplex networks 
which incorporates the process of triadic closure, the MTCVM. This process has been shown to be crucial 
in modeling the structure formation of individual layers in the {\em Pardus} society \cite{Thurner2013}.
We studied the phase diagram of the new model and found that it exhibits an anomalous fragmentation 
transition for multiplex networks that makes the model interesting in its own right.
This transition is characterized by a break-up of the largest component of a network layer into a large number of small components.
Intriguingly, the crucial parameters of the model turn out to be the differences of time-scales on which the link re-wiring 
dynamics takes place in the individual layers. 
When the model is calibrated to mimic the {\em Pardus} data on two different two-layer multiplex 
networks, community size distributions are perfectly compatible with those found in the data.

In particular the model confirms that {\em slow} layers in terms of the time-span between two re-wiring events, 
show a power-law-like distribution of community sizes, whereas the {\em fast} layers display an additional peak around community sizes of 50.
This means that the empirical community structure of the {\em Pardus} virtual society indeed resembles the fragmentation behavior predicted by the MTCVM.
Note that for these results the model layers only differ in their re-wiring rates (but not, for instance, in their degree),
 so that these results can only be attributed to the multiplex interaction of dynamics on different time-scales.
We further confirmed for an extended and calibrated version of the MTCVM that node- and link-based similarities 
between two pairs of layers in the data are indeed in good agreement with results from the model. 
These results suggest that the dynamical origin of the community structure of societies can be understood 
through the interplay of triadic closure processes taking place on different time-scales, which manifests itself in 
the phenomenon of shattered fragmentation.
Whether these results hold for empirical data in real-world societies remains to be seen. 

{\it Acknowledgments:}
We acknowledge financial support from FP7 projects LASAGNE, agreement no. 318132, MULTIPLEX, agreement no. 317532, and the Austrian Science Fund FWF, no. KPP23378FW. 

\bibliography{CVM_Bibliography}

\pagebreak

\section{Supplementary Information}

\subsubsection{Further Results on the Co-evolving Voter Model with Triadic Closure (TCVM)}

{\bf Triadic closure in the monoplex.}
A finite monoplex with standard, unlimited rewiring range ($p_{tc} = 0$) always reaches an absorbing state. This holds for the TCVM model except for the extreme $p = 1, p_{tc} = 1$ case. Here, limiting the rewiring range introduces dynamical traps, where `active' links between nodes in differing states continue to exist as they cannot be changed. This happens when the active link between $i$ and $j$ is also both $i^{th}$ and $j^{th}$ \emph{only} active link. An isolated node pair in differing states is one such example. Clearly as long as $p_{tc} < 1$ or $p = 1$ this trap is avoided as either one of the nodes rewires to someone further away, or eventually one of the nodes changes states. Existence of dynamical traps is just one of the consequences of a limited rewiring range. Another implication is that once a node is isolated, it cannot be brought back into a component, and hence the number of isolated components is non-decreasing.

\begin{figure}
\centering
\includegraphics[width = 0.49\textwidth, keepaspectratio = true]{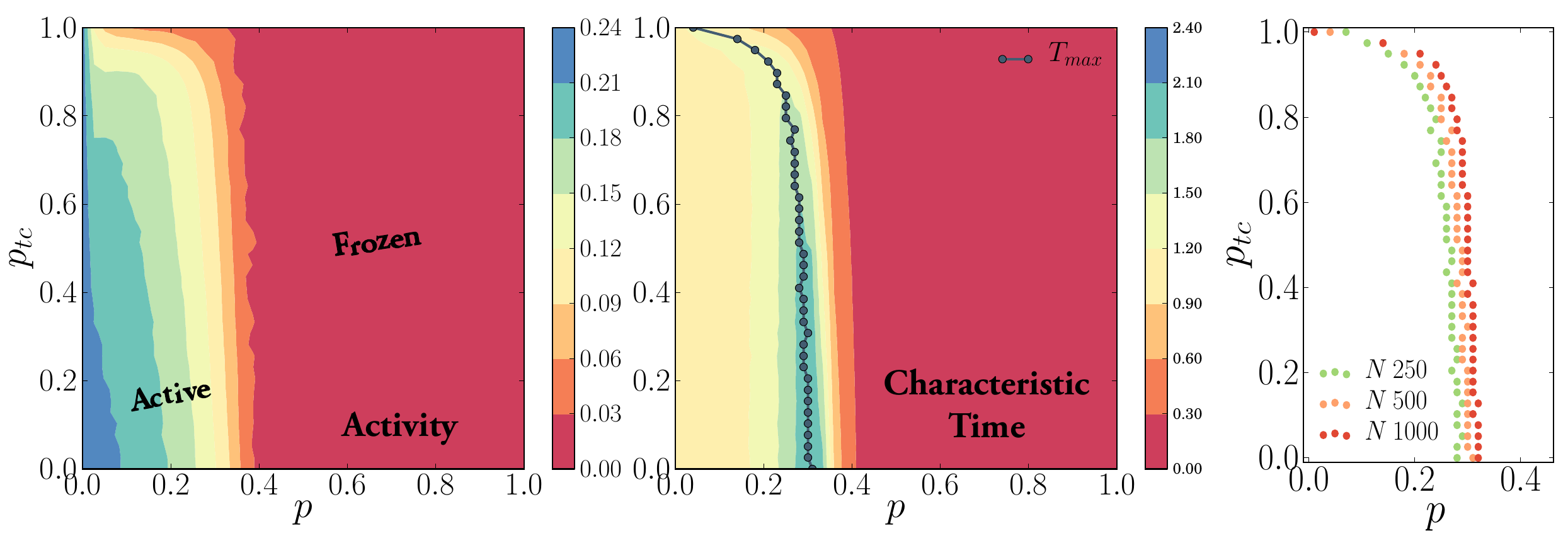}
\caption[]{Activity in the TCVM monoplex. Activity (left panel) as measured in the asymptotic value of the interface density averaged over surviving realizations in an ensemble of $10^3$ elements, for a network with $N = 500$. The characteristic time, i.e. the average time until absorption, for the same system (center panel), normalized by system size. The maximum (traced in dark circles) is associated with the fragmentation transition. Right panel: The variation of the maximum of characteristic time for various system sizes, computed from ensembles with $10^4$ elements.}
\label{TC_CVM_Activity}
\end{figure}

\begin{figure}
\centering
\includegraphics[width = 0.49\textwidth, keepaspectratio = true]{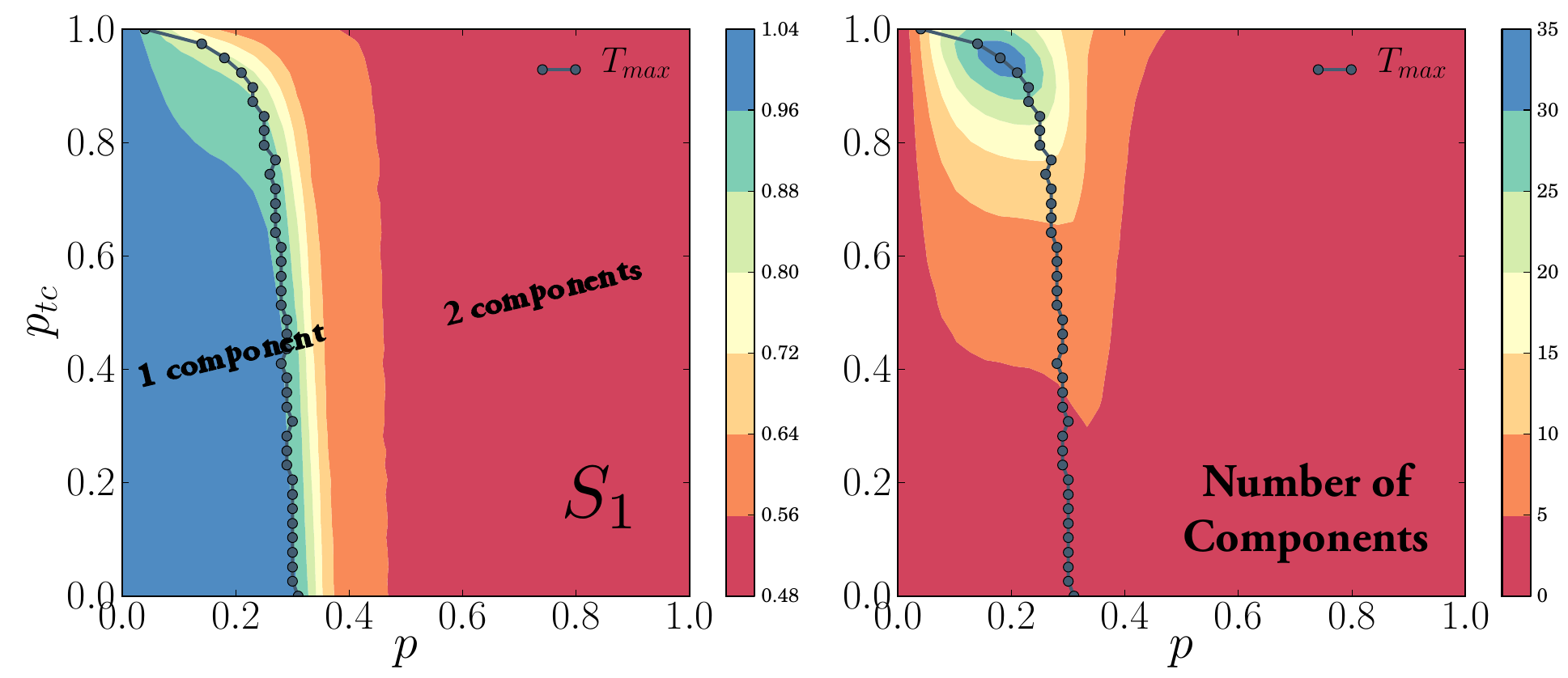}
\caption[]{Fragmentation in the TCVM monoplex. Each point is an average over absorbing configurations of $10^3$ realizations of the network with $N = 500$. Left panel: Fraction of nodes in the largest component, $S_1$; Right panel: the number of components. The dotted line denoted $T_{max}$ gives an indication of the absorbing/fragmentation transition.}
\label{TC_CVM_Fragmentation}
\end{figure}

Activity in the model is measured by the interface density, i.e. the fraction of active links which are the ones connecting nodes in different states. Figure~\ref{TC_CVM_Activity} shows the activity of the system in terms of asymptotic of the interface density averaged over surviving runs.  When the interface density is zero the system is in an absorbing state. The $p_{tc} = 0$ limit corresponds to the CVM (\cite{Vazquez2008}), where activity decreases with rewiring and falls to zero around $p_c$, defined in the limit of infinite systems. This critical rewiring denotes the absorbing transition and is accompanied by the critical slowing down of the time it takes for the system to reach an absorbing state. The average of such times is identified with the characteristic time, plotted in the middle panel of fig.~\ref{TC_CVM_Activity}. We associate its peak, $T_{\text{max}}$, with the finite-size approximation of the absorbing transition. Fig.~\ref{TC_CVM_Activity} shows that increasing the fraction of triadic closure decreases activity, and brings down the critical rewiring $p_c$. In other words, limiting the range of rewiring causes systems to freeze for an even lower rewiring range. The right panel of fig.~\ref{TC_CVM_Activity}, which traces the effect of system size of $p_c$, suggests that the absorbing transition does not exist when $p_{tc} = 1$; large systems rewiring with \emph{only} triadic closure do not sustain a constant level of activity, and will always reach an absorbing state in finite time.

The corresponding fragmentation diagram, overlayed by the trend of $T_{\text{max}}$ computed earlier, (fig.~\ref{TC_CVM_Fragmentation}) shows that for the vast majority of parameters in the TCVM the absorbing and the fragmentation transitions once more coincide: active systems ($p < p_c$) will reach an absorbing state with one giant component with all nodes in the same state, and frozen systems ($p > p_c$) will split into two components corresponding to the two initially present states (this is corroborated by examining the size of the second largest component, figure not shown). The novel aspect is the different nature of fragmentation observed for large values of $p_{tc}$, and maximized at $p_c$. There the decrease in the size of the largest component is no longer compensated only by the growth of the second largest component, but is accompanied by an increase in the number of small components that tend to be isolated nodes or pairs of isolated nodes. This is an example of shattered fragmentation: topologies with one or two giant components and a multiplicity of components of negligible size. This was first observed in the CVM on a multiplex  \cite{Diakonova2014} for incomplete interlayer connectivity $q < 1$, and in \cite{DiakonovaNoise} for the CVM on a monoplex with noise. Here we see that a similar effect can be caused by limiting the scope of rewiring.


\end{document}